# Communication Lower Bounds Using Dual Polynomials[*]

ALEXANDER A. SHERSTOV
*Univ. of Texas at Austin, Dept. of Comp. Sciences, sherstov@cs.utexas.edu*

November 9, 2018

**Abstract**

Representations of Boolean functions by real polynomials play an important role in complexity theory. Typically, one is interested in the least degree of a polynomial $p(x_1, \ldots, x_n)$ that approximates or sign-represents a given Boolean function $f(x_1, \ldots, x_n)$. This article surveys a new and growing body of work in communication complexity that centers around the *dual* objects, i.e., polynomials that certify the difficulty of approximating or sign-representing a given function. We provide a unified guide to the following results, complete with all the key proofs:

- Sherstov's *Degree/Discrepancy Theorem,* which translates lower bounds on the threshold degree of a Boolean function into upper bounds on the discrepancy of a related function;

- Two different methods for proving lower bounds on bounded-error communication based on the approximate degree: Sherstov's *pattern matrix method* and Shi and Zhu's *block composition method*;

- Extension of the pattern matrix method to the multiparty model, obtained by Lee and Shraibman and by Chattopadhyay and Ada, and the resulting improved lower bounds for DISJOINTNESS;

- David and Pitassi's separation of NP and BPP in multiparty communication complexity for $k \leq (1 - \epsilon) \log n$ players.

---

# 1 Introduction

Representations of Boolean functions by real polynomials are of considerable importance in complexity theory. The ease or difficulty of representing a given Boolean function by polynomials from a given set often yields valuable insights into the structural complexity of that function.

We focus on two concrete representation schemes that involve polynomials. The first of these corresponds to threshold computation. For a Boolean function $f : \{0, 1\}^n \to \{0, 1\}$, its *threshold degree* $\deg_{\pm}(f)$ is the minimum degree of a polynomial $p(x_1, \ldots, x_n)$ such that $p(x)$ is positive if $f(x) = 1$ and negative otherwise. In other words, the threshold degree of $f$ is the least degree of a polynomial that represents $f$ in sign. Several authors have analyzed the threshold degree of common Boolean functions [MP88, BRS95, OS03]. The results of these investigations have found numerous applications to circuit complexity [ABFR94, BRS95, KP97, KP98] and computational learning theory [KS04, KOS04, KS07b].

The other representation scheme that we consider is approximation in the uniform norm. For a Boolean function $f : \{0, 1\}^n \to \{0, 1\}$ and a constant $\epsilon \in (0, 1/2)$, the $\epsilon$-approximate degree of $f$ is the least degree of a polynomial $p(x_1, \ldots, x_n)$ with $|f(x) - p(x)| \leq \epsilon$ for all $x \in \{0, 1\}^n$. Note that this representation is strictly stronger than the first: no longer are we content with representing $f$ in sign, but rather we wish to closely approximate $f$ on every input. There is a considerable literature on the approximate degree of specific Boolean functions [NS92, Pat92, KLS96, BCWZ99, AS04, She08, Wol08]. This classical notion has been crucial to progress on a variety of questions, including quantum query complexity [BCWZ99, BBC+01, AS04], communication complexity [BW01, Raz03, BVW07] and computational learning theory [TT99, KS04, KKMS05, KS07a].

The approximate degree and threshold degree can be conveniently analyzed by means of a linear program. In particular, whenever a given function $f$ cannot be approximated or sign-represented by polynomials of low degree, linear-programming duality implies the existence of a certain *dual* object to witness that fact. This dual object, which is a real function or a probability distribution, reveals useful new information about the structural complexity of $f$. The purpose of this article is to survey a very recent and growing body of work in communication complexity that revolves around the dual formulations of the approximate degree and threshold degree. Our ambition here is to provide a unified view of these diverse results, complete with all the key proofs, and thereby to encourage further inquiry into the potential of the dual approach.

In the remainder of this section, we give an intuitive overview of our survey.

**Degree/Discrepancy Theorem.** The first result that we survey, in Section 3, is the author's *Degree/Discrepancy Theorem* [She07a]. This theorem and its proof technique are the foundation for much of the subsequent work surveyed in this article [She07b, Cha07, LS07, CA08, DP08]. Fix a Boolean function $f : \{0, 1\}^n \to \{0, 1\}$ and let $N$ be a given integer, $N \geq n$. In [She07a], we introduced the two-party communication problem of computing

$$f(x|_V),$$

where the Boolean string $x \in \{0, 1\}^N$ is Alice's input and the set $V \subset \{1, 2, \ldots, N\}$ of size $|V| = n$ is Bob's input. The symbol $x|_V$ stands for the projection of $x$ onto the indices in $V$, in other words, $x|_V = (x_{i_1}, x_{i_2}, \ldots, x_{i_n}) \in \{0, 1\}^n$, where $i_1 < i_2 < \cdots < i_n$ are the elements of $V$. Intuitively, this problem models a situation when Alice and Bob's joint computation depends on only $n$ of the inputs $x_1, x_2, \ldots, x_N$. Alice knows the values of all the inputs $x_1, x_2, \ldots, x_N$ but does not know which $n$ of them are relevant. Bob, on the other hand, knows which $n$ inputs are relevant but does not know their values.

We proved in [She07a] that the threshold degree $d$ of $f$ is a lower bound on the communication requirements of this problem. More precisely, the Degree/Discrepancy Theorem shows that this communication problem has discrepancy $\exp(-\Omega(d))$ as soon as $N \geq 11n^2/d$. This exponentially small discrepancy immediately gives an $\Omega(d)$ lower bound on communication in a variety of models (deterministic, nondeterministic, randomized, quantum with and without entanglement). Moreover, the resulting lower bounds on communication hold even if the desired error probability is vanishingly close to $1/2$.

The proof of the Degree/Discrepancy Theorem introduces a novel technique based on the dual formulation of the threshold degree. In fact, it appears to be the first use of the threshold degree (in its primal or dual form) to prove communication lower bounds. As an application, we exhibit in [She07a] the first $\mathsf{AC}^0$ circuit with exponentially small discrepancy, thereby separating $\mathsf{AC}^0$ from depth-2 majority circuits and solving an open problem of Krause and Pudlák [KP97, §6]. Independently of the author, Buhrman et al. [BVW07] exhibited another $\mathsf{AC}^0$ function with exponentially small discrepancy, using much different techniques.

**Bounded-Error Communication.** Next, we present two recent results on bounded-error communication complexity, due to Sherstov [She07b] and Shi and Zhu [SZ07]. These papers use the notion of approximate degree to contribute strong lower bounds for rather broad classes of functions, subsuming Razborov's breakthrough work on symmetric predicates [Raz03]. The lower bounds are valid not only in the randomized model, but also in the quantum model with and without prior entanglement.

The setting in which to view these two works is the *generalized discrepancy method,* a simple but very useful principle introduced by Klauck [Kla01] and reformulated in its current form by Razborov [Raz03]. Let $f(x, y)$ be a Boolean function whose quantum communication complexity is of interest. The method asks for a Boolean function $h(x, y)$ and a distribution $\mu$ on $(x, y)$-pairs such that:

(1) the functions $f$ and $h$ are highly correlated under $\mu$; and

(2) all low-cost protocols have negligible advantage in computing $h$ under $\mu$.

If such $h$ and $\mu$ indeed exist, it follows that no low-cost protocol can compute $f$ to high accuracy (or else it would be a good predictor for the hard function $h$ as well!). This method is in no way restricted to the quantum model but, rather, applies to *any* model of communication [She07b, §2.4]. The importance of the generalized discrepancy method is that it makes it possible, in theory, to prove lower bounds for functions such as DISJOINTNESS, to which the traditional discrepancy method does not apply. In Section 4, we provide detailed historical background on the generalized discrepancy method and compile its quantitative versions for several models.

The hard part, of course, is finding $h$ and $\mu$. Except in rather restricted cases [Kla01, Thm. 4], it was not known how to do it. As a result, the generalized discrepancy method was of limited practical use. This difficulty was overcome independently by Sherstov [She07b] and Shi and Zhu [SZ07], who used the dual characterization of the approximate degree to obtain $h$ and $\mu$ for a broad range of problems. To our knowledge, the work in [She07b] and [SZ07] is the first use of the dual characterization of the approximate degree to prove communication lower bounds. The specifics of these two works are very different. The construction of $h$ and $\mu$ in [She07b], which we called the *pattern matrix method* for lower bounds on bounded-error communication, is built around a new matrix-analytic technique (the *pattern matrix*) inspired by the author's Degree/Discrepancy Theorem. The construction in [SZ07], the *block-composition method,* is based on the idea of hardness amplification by composition. These two methods exhibit quite different behavior, e.g., the pattern matrix method further extends to the multiparty model. We present the two methods individually in Sections 5.1 and 5.2 and provide a detailed comparison of their strength and applicability in Section 5.3.

**Extensions to the Multiparty Model.** Both the Degree/Discrepancy Theorem [She07a] and the pattern matrix method [She07b] generalize to the multiparty number-on-the-forehead model. In the case of [She07a], this extension was formalized by Chattopadhyay [Cha07]. As before, let $f : \{0, 1\}^n \to \{0, 1\}$ be a given function. Recall that in the *two-party* case, there was a Boolean string $x \in \{0, 1\}^N$ and a *single* set $V \subset \{1, 2, \ldots, N\}$. The *k-party* communication problem features a Boolean string $x \in \{0, 1\}^{N^{k-1}}$ and sets $V_1, \ldots, V_{k-1} \subset \{1, 2, \ldots, N\}$. The $k$ inputs $x, V_1, \ldots, V_{k-1}$ are distributed among the $k$ parties as usual. The goal is to compute

$$f(x|_{V_1, \ldots, V_{k-1}}) \stackrel{\text{def}}{=} f\left(x_{i_1^1, \ldots, i_1^{k-1}}, \ldots, x_{i_n^1, \ldots, i_n^{k-1}}\right), \tag{1.1}$$

where $i_1^j < i_2^j < \cdots < i_n^j$ are the elements of $V_j$ (for $j = 1, 2, \ldots, k-1$). This way, again no party knows at once the Boolean string $x$ and the relevant bits in it. With this setup in place, it becomes relatively straightforward to bound the discrepancy by traversing the same line of reasoning as in [She07a]. The extension of the pattern matrix method [She07b] to the multiparty model uses a similar setup and was done by Lee and Shraibman [LS07] and independently by Chattopadhyay and Ada [CA08]. We present the proofs of these extensions in Section 6, placing them in close correspondence with the two-party case. These extensions do not subsume the two-party results, however (see Section 6 for details).

The authors of [LS07] and [CA08] gave important applications of their work to the $k$-party randomized communication complexity of DISJOINTNESS, improving it from $\Omega(\frac{1}{k} \log n)$ to $n^{\Omega(1/k)} 2^{-O(2^k)}$. As a corollary, they separated the multiparty communication classes $\mathsf{NP}_k^{cc}$ and $\mathsf{BPP}_k^{cc}$ for $k = (1 - o(1)) \log_2 \log_2 n$ parties. They also obtained new results for Lovász-Schrijver proof systems, in light of the work due to Beame, Pitassi, and Segerlind [BPS07].

**Separation of $\mathsf{NP}_k^{cc}$ and $\mathsf{BPP}_k^{cc}$.** The separation of the classes $\mathsf{NP}_k^{cc}$ and $\mathsf{BPP}_k^{cc}$ in [LS07, CA08] for $k = (1 - o(1)) \log_2 \log_2 n$ parties was followed by another exciting development, due to David and Pitassi [DP08], who separated these classes for $k \leq (1 - \epsilon) \log_2 n$ parties. Here $\epsilon > 0$ is an arbitrary constant. Since the current barrier for explicit lower bounds on multiparty communication complexity is precisely $k = \log_2 n$, David and Pitassi's separation matches the state of the art. We present this work in Section 7. The powerful idea in this result was to redefine the projection operator $x|_{V_1, \ldots, V_{k-1}}$ in (1.1). Specifically, David and Pitassi observed that it suffices to define the projection operator at random, using the probabilistic method. This insight removed the key technical obstacle present in [LS07, CA08]. In a follow-up work by David, Pitassi, and Viola [DPV08], the probabilistic construction was derandomized to yield an *explicit* separation.

**Other Related Work.** For completeness, we will mention several duality-based results in communication complexity that fall outside the scope of this survey. Recent work has seen other applications of dual polynomials [She07c, RS08], which are considerably more complicated and no longer correspond to the approximate degree or threshold degree. More broadly, several recent results feature other forms of duality [LS07b, LSŠ08], such as the duality of norms or semidefinite programming duality.

## 2 Preliminaries

This section reviews our notation and provides relevant technical background.

### 2.1 General Background

A *Boolean function* is a mapping $X \to \{0, 1\}$, where $X$ is a finite set such as $X = \{0, 1\}^n$ or $X = \{0, 1\}^n \times \{0, 1\}^n$. The notation $[n]$ stands for the set $\{1, 2, \ldots, n\}$. For integers $N, n$ with $N \geq n$, the symbol $\binom{[N]}{n}$ denotes the family of all size-$n$ subsets of $\{1, 2, \ldots, N\}$. For $x \in \{0, 1\}^n$, we write $|x| = x_1 + \cdots + x_n$. For $x, y \in \{0, 1\}^n$, the notation $x \wedge y$ refers as usual to the component-wise AND of $x$ and $y$. In particular, $|x \wedge y|$ stands for the number of positions where $x$ and $y$ both have a 1. Throughout this manuscript, "log" refers to the logarithm to base 2.

For tensors $A, B : X_1 \times \cdots \times X_k \to \mathbb{R}$ (where $X_i$ is a finite set, $i = 1, 2, \ldots, k$), define $\langle A, B \rangle = \sum_{(x_1, \ldots, x_k) \in X_1 \times \cdots \times X_k} A(x_1, \ldots, x_k) B(x_1, \ldots, x_k)$. When $A$ and $B$ are vectors or matrices, this is the standard definition of inner product. The *Hadamard product* of $A$ and $B$ is the tensor $A \circ B : X_1 \times \cdots \times X_k \to \mathbb{R}$ given by $(A \circ B)(x_1, \ldots, x_k) = A(x_1, \ldots, x_k) B(x_1, \ldots, x_k)$.

The symbol $\mathbb{R}^{m \times n}$ refers to the family of all $m \times n$ matrices with real entries. The $(i, j)$th entry of a matrix $A$ is denoted by $A_{ij}$. We frequently use "generic-entry" notation to specify a matrix succinctly: we write $A = [F(i, j)]_{i,j}$ to mean that the $(i, j)$th entry of $A$ is given by the expression $F(i, j)$. In most matrices that arise in this work, the exact ordering of the columns (and rows) is irrelevant. In such cases we describe a matrix by the notation $[F(i, j)]_{i \in I, j \in J}$, where $I$ and $J$ are some index sets.

Let $A \in \mathbb{R}^{m \times n}$. We use the following standard notation: $\|A\|_\infty = \max_{i,j} |A_{ij}|$ and $\|A\|_1 = \sum_{i,j} |A_{ij}|$. We denote the singular values of $A$ by $\sigma_1(A) \geq \sigma_2(A) \geq \ldots \geq \sigma_{\min\{m,n\}}(A) \geq 0$. Recall that the spectral norm of $A$ is given by $\|A\| = \max_{x \in \mathbb{R}^n, \|x\|=1} \|Ax\| = \sigma_1(A)$. An excellent reference on matrix analysis is [HJ86].

We conclude with a review of the Fourier transform over $\mathbb{Z}_2^n$. Consider the vector space of functions $\{0, 1\}^n \to \mathbb{R}$, equipped with the inner product $\langle f, g \rangle = 2^{-n} \sum_{x \in \{0,1\}^n} f(x) g(x)$. For $S \subseteq [n]$, define $\chi_S : \{0, 1\}^n \to \{-1, +1\}$ by $\chi_S(x) = (-1)^{\sum_{i \in S} x_i}$. Then $\{\chi_S\}_{S \subseteq [n]}$ is an orthonormal basis for the inner product space in question. As a result, every function $f : \{0, 1\}^n \to \mathbb{R}$ has a unique representation of the form $f(x) = \sum_{S \subseteq [n]} \hat{f}(S) \chi_S(x)$, where $\hat{f}(S) = \langle f, \chi_S \rangle$. The reals $\hat{f}(S)$ are called the *Fourier coefficients of f*. The following fact is immediate from the definition of $\hat{f}(S)$:

**Proposition 2.1.** *Fix* $f : \{0,1\}^n \to \mathbb{R}$. *Then*

$$\max_{S \subseteq [n]} |\hat{f}(S)| \leq 2^{-n} \sum_{x \in \{0,1\}^n} |f(x)|.$$

## 2.2 Communication Complexity

This survey features several standard models of communication. In the case of two communicating parties, one considers a function $f : X \times Y \to \{0, 1\}$, where $X$ and $Y$ are some finite sets. Alice receives an input $x \in X$, Bob receives $y \in Y$, and their objective is to predict $f(x, y)$ with good accuracy. To this end, Alice and Bob share a communication channel (classical or quantum, depending on the model). Alice and Bob's communication protocol is said to have *error* $\epsilon$ if it outputs the correct answer $f(x, y)$ with probability at least $1 - \epsilon$ on every input. The *cost* of a given protocol is the maximum number of bits exchanged on any input. The two-party models of interest to us are the randomized model, the quantum model without prior entanglement, and the quantum model with prior entanglement. The least cost of an $\epsilon$-error protocol for $f$ in these models is denoted by $R_\epsilon(f)$, $Q_\epsilon(f)$, and $Q^*_\epsilon(f)$, respectively. It is standard practice to omit the subscript $\epsilon$ when error parameter is $\epsilon = 1/3$. Recall that the error probability of a protocol can be decreased from $1/3$ to any other constant $\epsilon > 0$ at the expense of increasing the communication cost by a constant factor; we will use this fact in many proofs of this survey, often without explicitly mentioning it. Excellent references on these communication models are [KN97] and [Wol01].

A generalization of two-party communication is *number-on-the-forehead multiparty* communication. Here one considers a function $f : X_1 \times \cdots \times X_k \to \{0, 1\}$ for some finite sets $X_1, \ldots, X_k$. There are $k$ players. A given input $(x_1, \ldots, x_k) \in X_1 \times \cdots \times X_k$ is distributed among the players by placing $x_i$ on the forehead of player $i$ (for $i = 1, \ldots, k$). In other words, player $i$ knows $x_1, \ldots, x_{i-1}, x_{i+1}, \ldots, x_k$ but not $x_i$. The players can communicate by writing bits on a shared blackboard, visible to all. They additionally have access to a shared source of random bits. Their goal is to devise a communication protocol that will allow them to accurately predict the value of $f$ on every input. Analogous to the two-party case, the *randomized* communication complexity $R^k_\epsilon(f)$ is the least cost of an $\epsilon$-error communication protocol for $f$ in this model. The final section of this paper also considers the *nondeterministic* communication complexity $N^k(f)$, which is the minimum cost of a protocol for $f$ that always outputs the correct answer on the inputs $f^{-1}(0)$ and has error probability less than 1 on each of the inputs $f^{-1}(1)$. Analogous to computational complexity, $\mathsf{BPP}_k^{cc}$ (respectively, $\mathsf{NP}_k^{cc}$) is the class of functions $f : (\{0,1\}^n)^k \to \{0,1\}$ with $R^k(f) \leq (\log n)^{O(1)}$ (respectively, $N^k(f) \leq (\log n)^{O(1)}$). See [KN97] for further details.

A crucial tool for proving communication lower bounds is the *discrepancy method.* Given a function $f : X \times Y \to \{0,1\}$ and a distribution $\mu$ on $X \times Y$, the *discrepancy of $f$ with respect to $\mu$* is defined as

$$\mathrm{disc}_\mu(f) = \max_{\substack{S \subseteq X, \\ T \subseteq Y}} \left| \sum_{x \in S} \sum_{y \in T} (-1)^{f(x,y)} \mu(x,y) \right|.$$

This definition generalizes to the multiparty case as follows. Fix $f : X_1 \times \cdots \times X_k \to \{0,1\}$ and a distribution $\mu$ on $X_1 \times \cdots \times X_k$. The *discrepancy of $f$ with respect to $\mu$* is defined as

$$\mathrm{disc}_\mu(f) = \max_{\phi_1,\ldots,\phi_k} \left| \sum_{\substack{(x_1,\ldots,x_k) \\ \in X_1 \times \cdots \times X_k}} \psi(x_1,\ldots,x_k) \prod_{i=1}^k \phi_i(x_1,\ldots,x_{i-1},x_{i+1},\ldots,x_k) \right|,$$

where $\psi(x_1,\ldots,x_k) = (-1)^{f(x_1,\ldots,x_k)} \mu(x_1,\ldots,x_k)$ and the maximum ranges over all functions $\phi_i : X_1 \times \cdots X_{i-1} \times X_{i+1} \times \cdots X_k \to \{0,1\}$, for $i = 1, 2, \ldots, k$. Note that for $k = 2$, this definition is identical to the one given previously for the two-party model. We put $\mathrm{disc}(f) = \min_\mu \mathrm{disc}_\mu(f)$. We identify a function $f : X_1 \times \cdots \times X_k \to \{0,1\}$ with its *communication tensor* $M(x_1,\ldots,x_k) = (-1)^{f(x_1,\ldots,x_k)}$ and speak of the discrepancy of $M$ and $f$ interchangeably (and likewise for other complexity measures, such as $R^k(f)$).

Discrepancy is difficult to analyze as defined. Typically, one uses the following well-known estimate, derived by repeated applications of the Cauchy-Schwartz inequality.

**Theorem 2.2** ([BNS92, CT93, Raz00]). *Fix $f : X_1 \times \cdots \times X_k \to \{0,1\}$ and a distribution $\mu$ on $X_1 \times \cdots \times X_k$. Put $\psi(x_1,\ldots,x_k) = (-1)^{f(x_1,\ldots,x_k)} \mu(x_1,\ldots,x_k)$. Then*

$$\left( \frac{\mathrm{disc}_\mu(f)}{|X_1| \cdots |X_k|} \right)^{2^{k-1}} \leq \mathop{\mathbf{E}}_{\substack{x_1^0 \in X_1 \\ x_1^1 \in X_1}} \cdots \mathop{\mathbf{E}}_{\substack{x_{k-1}^0 \in X_{k-1} \\ x_{k-1}^1 \in X_{k-1}}} \left| \mathop{\mathbf{E}}_{x_k \in X_k} \prod_{z \in \{0,1\}^{k-1}} \psi(x_1^{z_1},\ldots,x_{k-1}^{z_{k-1}},x_k) \right|.$$

In the case of $k = 2$ parties, there are other ways to estimate the discrepancy, e.g., using the spectral norm of a matrix.

For a function $f : X_1 \times \cdots \times X_k \to \{0, 1\}$ and a distribution $\mu$ over $X_1 \times \cdots \times X_k$, let $D_\epsilon^{k,\mu}(f)$ denote the least cost of a deterministic protocol for $f$ whose probability of error with respect to $\mu$ is at most $\epsilon$. This quantity is known as the *$\mu$-distributional complexity* of $f$. Since a randomized protocol can be viewed as a probability distribution over deterministic protocols, we immediately have that $R_\epsilon^k(f) \geq \max_\mu D_\epsilon^{k,\mu}(f)$. We are now ready to state the discrepancy method.

**Theorem 2.3** (Discrepancy method; see [KN97]). *For every $f : X_1 \times \cdots \times X_k \to \{0, 1\}$, every distribution $\mu$ on $X_1 \times \cdots \times X_k$, and every $\gamma \in (0, 1]$,*

$$R_{1/2-\gamma/2}^k \geq D_{1/2-\gamma/2}^{k,\mu}(f) \geq \log_2 \frac{\gamma}{\mathrm{disc}_\mu(f)}.$$

In other words, a function with small discrepancy is hard to compute to any nontrivial advantage over random guessing (let alone compute it to high accuracy). In the case of $k = 2$ parties, discrepancy yields analogous lower bounds even in the quantum model, regardless of prior entanglement [Kre95, Kla01, LS07b].

## 3 The Degree/Discrepancy Theorem

This section presents the author's Degree/Discrepancy Theorem, whose proof technique is the foundation for much of the subsequent work surveyed in this article [She07b, Cha07, LS07, CA08, DP08].

The original motivation behind this result came from circuit complexity. A natural and well-studied computational model is that of a polynomial-size circuit of majority gates. Research has shown that majority circuits of depth 2 and 3 already possess surprising computational power. Indeed, it is a long-standing open problem [KP97] to exhibit a Boolean function that *cannot* be computed by a depth-3 majority circuit of polynomial size.

Another extensively studied model is that of polynomial-size constant-depth circuits with AND, OR, NOT gates, denoted by $\mathsf{AC}^0$. Allender's classic result [All89] states that every function in $\mathsf{AC}^0$ can be computed by a depth-3 majority circuit of quasipolynomial size. Krause and Pudlák [KP97, §6] ask whether this simulation can be improved, i.e., whether every function in $\mathsf{AC}^0$ can be computed by a depth-2 majority circuit of quasipolynomial size. We recently gave a strong negative answer to this question:

**Theorem 3.1** ([She07a]). *There is a function $F : \{0, 1\}^n \to \{0, 1\}$, explicitly given and computable by an $\mathsf{AC}^0$ circuit of depth $3$, whose computation requires a majority vote of $\exp(\Omega(n^{1/5}))$ threshold gates.*

We proved Theorem 3.1 by exhibiting an $\mathsf{AC}^0$ function with exponentially small discrepancy. All previously known functions with exponentially small discrepancy (e.g., [GHR92, Nis93]) contained PARITY or MAJORITY as a subfunction and therefore could not be computed in $\mathsf{AC}^0$. Buhrman et al. [BVW07] obtained, independently of the author and with much different techniques, another $\mathsf{AC}^0$ function with exponentially small discrepancy, thereby also answering Krause and Pudlák's question.

## 3.1 Bounding the Discrepancy via the Threshold Degree

To construct an $\mathsf{AC}^0$ function with small discrepancy, we developed in [She07a] a novel technique for generating low-discrepancy functions, which we now describe. This technique is not specialized in any way to $\mathsf{AC}^0$ but, rather, is based on the abstract notion of threshold degree.

For a Boolean function $f : \{0, 1\}^n \to \{0, 1\}$, recall from Section 1 that its *threshold degree* $\deg_\pm(f)$ is the minimum degree of a polynomial $p(x_1, \ldots, x_n)$ with $p(x) > 0 \Leftrightarrow f(x) = 1$ and $p(x) < 0 \Leftrightarrow f(x) = 0$. In many cases [MP88], it is straightforward to obtain strong lower bounds on the threshold degree. Since the threshold degree is a measure of the complexity of a given Boolean function, it is natural to wonder whether it can yield lower bounds on communication in a suitable setting. As we prove in [She07a], this intuition turns out to be correct for every $f$.

More precisely, fix a Boolean function $f : \{0, 1\}^n \to \{0, 1\}$ with threshold degree $d$. Let $N$ be a given integer, $N \geqslant n$. In [She07a], we introduced the two-party communication problem of computing

$$f(x|_V),$$

where the Boolean string $x \in \{0, 1\}^N$ is Alice's input and the set $V \subset \{1, 2, \ldots, N\}$ of size $|V| = n$ is Bob's input. The symbol $x|_V$ stands for the projection of $x$ onto the indices in $V$, in other words, $x|_V = (x_{i_1}, x_{i_2}, \ldots, x_{i_n}) \in \{0, 1\}^n$, where $i_1 < i_2 < \cdots < i_n$ are the elements of $V$. Intuitively, this problem models a situation when Alice and Bob's joint computation depends on only $n$ of the inputs $x_1, x_2, \ldots, x_N$. Alice knows the values of all the inputs $x_1, x_2, \ldots, x_N$ but does not know which $n$ of them are relevant. Bob, on the other hand, knows which $n$ inputs are relevant but does not know their values. As one would hope, it turns out that $d$ is a lower bound on the communication requirements of this problem:

**Theorem 3.2** (Degree/Discrepancy Theorem [She07a]). *Let $f : \{0,1\}^n \to \{0,1\}$ be given with threshold degree $d \geq 1$. Let $N$ be a given integer, $N \geq n$. Define $F = [f(x|_V)]_{x,V}$, where the rows are indexed by $x \in \{0,1\}^N$ and columns by $V \in \binom{[N]}{n}$. Then*

$$\mathrm{disc}(F) \leq \left(\frac{4en^2}{Nd}\right)^{d/2}.$$

To our knowledge, Theorem 3.2 is the first use of the threshold degree to prove communication lower bounds. Given a function $f$ with threshold degree $d$, Theorem 3.2 generates a communication problem with discrepancy at most $2^{-d}$ (by setting $N \geq 16en^2/d$). This exponentially small discrepancy immediately gives an $\Omega(d)$ lower bound on communication in a variety of models (deterministic, nondeterministic, randomized, quantum with and without entanglement; see Section 2.2). Moreover, the resulting lower bounds on communication remain valid when Alice and Bob merely seek to predict the answer with nonnegligible advantage, a critical aspect for lower bounds against threshold circuits.

We will give a detailed proof of the Degree/Discrepancy Theorem in the next subsection. For now we will briefly sketch how we used it in [She07a] to prove the main result of that paper, Theorem 3.1 above, on the existence of an $\mathsf{AC}^0$ function that requires a depth-2 majority circuit of exponential size. Consider the function

$$f(x) = \bigvee_{i=1}^{m} \bigwedge_{j=1}^{4m^2} x_{ij},$$

for which Minsky and Papert [MP88] showed that $\deg_{\pm}(f) = m$. Since $f$ has high threshold degree, an application of Theorem 3.2 to $f$ yields a communication problem with low discrepancy. This communication problem itself can be viewed as an $\mathsf{AC}^0$ circuit of depth 3. Recalling that its discrepancy is exponentially small, we conclude that it cannot be computed by a depth-2 majority circuit of subexponential size.

## 3.2 Proof of the Degree/Discrepancy Theorem

A key ingredient in our proof is the following dual characterization of the threshold degree, which is a classical result known in greater generality as Gordan's Transposition Theorem [Sch98, §7.8]:

**Theorem 3.3.** *Let $f : \{0,1\}^n \to \{0,1\}$ be arbitrary, $d$ a nonnegative integer. Then exactly one of the following holds:* (1) *$f$ has threshold degree at most $d$;* (2) *there is a distribution $\mu$ over $\{0,1\}^n$ such that $\mathbf{E}_{x \sim \mu}[(-1)^{f(x)}\chi_S(x)] = 0$ for $|S| = 0, 1, \ldots, d$.*

Theorem 3.3 follows from linear-programming duality. We will also make the following simple observation.

**Observation 3.4.** *Let $\kappa(x)$ be a probability distribution on $\{0,1\}^r$. Fix $i_1, \ldots, i_r \in \{1, 2, \ldots, r\}$. Then $\sum_{x \in \{0,1\}^r} \kappa(x_{i_1}, \ldots, x_{i_r}) \leq 2^{r-|\{i_1, \ldots, i_r\}|}$, where $|\{i_1, \ldots, i_r\}|$ denotes the number of distinct integers among $i_1, \ldots, i_r$.*

We are now ready for the proof of the Degree/Discrepancy Theorem.

**Theorem 3.2** (Restated from p. 10)**.** *Let $f : \{0,1\}^n \to \{0,1\}$ be given with threshold degree $d \geq 1$. Let $N$ be a given integer, $N \geq n$. Define $F = [f(x|_V)]_{x,V}$, where the rows are indexed by $x \in \{0,1\}^N$ and columns by $V \in \binom{[N]}{n}$. Then*

$$\mathrm{disc}(F) \leq \left(\frac{4en^2}{Nd}\right)^{d/2}.$$

*Proof* [She07a]. Let $\mu$ be a probability distribution over $\{0,1\}^n$ with respect to which $\mathbf{E}_{z \sim \mu}[(-1)^{f(z)} p(z)] = 0$ for every real-valued function $p$ of $d-1$ or fewer of the variables $z_1, \ldots, z_n$. The existence of $\mu$ is assured by Theorem 3.3. We will analyze the discrepancy of $F$ with respect to the distribution

$$\lambda(x, V) = 2^{-N+n} \binom{N}{n}^{-1} \mu(x|_V).$$

Define $\psi : \{0,1\}^n \to \mathbb{R}$ by $\psi(z) = (-1)^{f(z)} \mu(z)$. By Theorem 2.2,

$$\mathrm{disc}_\lambda(F)^2 \leq 4^n \, \mathbf{E}_{V,W} |\Gamma(V, W)|, \tag{3.1}$$

where we put $\Gamma(V, W) = \mathbf{E}_x[\psi(x|_V)\psi(x|_W)]$. To analyze this expression, we prove two key claims.

**Claim 3.5.** *Assume that $|V \cap W| \leq d - 1$. Then $\Gamma(V, W) = 0$.*

*Proof.* The claim is immediate from the fact that the Fourier transform of $\psi$ is supported on characters of order $d$ and higher. For completeness, we will now give a more detailed and elementary explanation. Assume for notational convenience that $V = \{1, 2, \ldots, n\}$. Then

$$\Gamma(V, W) = \mathbf{E}_x[\mu(x_1, \ldots, x_n)(-1)^{f(x_1, \ldots, x_n)} \psi(x|_W)]$$

$$= \frac{1}{2^N} \sum_{x_1, \ldots, x_n} \mu(x_1, \ldots, x_n)(-1)^{f(x_1, \ldots, x_n)} \sum_{x_{n+1}, \ldots, x_N} \psi(x|_W)$$

$$= \frac{1}{2^N} \mathbf{E}_{(x_1, \ldots, x_n) \sim \mu}\left[(-1)^{f(x_1, \ldots, x_n)} \cdot \underbrace{\left(\sum_{x_{n+1}, \ldots, x_N} \psi(x|_W)\right)}_{*}\right].$$

Since $|V \cap W| \leq d - 1$, the starred expression is a real-valued function of at most $d - 1$ variables. The claim follows by the definition of $\mu$. □

**Claim 3.6.** *Assume that $|V \cap W| = i$. Then $|\Gamma(V, W)| \leq 2^{i-2n}$.*

*Proof.* The claim is immediate from Observation 3.4. For completeness, we will give a more detailed explanation. For notational convenience, assume that

$$V = \{1, 2, \ldots, n\},$$
$$W = \{1, 2, \ldots, i\} \cup \{n + 1, n + 2, \ldots, n + (n - i)\}.$$

We have:

$$|\Gamma(V, W)| \leq \mathbf{E}_x[|\psi(x|_V)\psi(x|_W)|]$$
$$= \mathop{\mathbf{E}}_{x_1,\ldots,x_{2n-i}}[\mu(x_1, \ldots, x_n)\mu(x_1, \ldots, x_i, x_{n+1}, \ldots, x_{2n-i})]$$
$$\leq \underbrace{\mathop{\mathbf{E}}_{x_1,\ldots,x_n}[\mu(x_1, \ldots, x_n)]}_{=2^{-n}} \cdot \max_{x_1,\ldots,x_i} \underbrace{\mathop{\mathbf{E}}_{x_{n+1},\ldots,x_{2n-i}}[\mu(x_1, \ldots, x_k, x_{n+1}, \ldots, x_{2n-i})]}_{\leq 2^{-(n-i)}}.$$

The bounds $2^{-n}$ and $2^{-(n-i)}$ follow because $\mu$ is a probability distribution. □

In view of Claims 3.5 and 3.6, inequality (3.1) simplifies to

$$\mathrm{disc}_\lambda(F)^2 \leq \sum_{i=d}^{n} 2^i \, \mathbf{P}[|V \cap W| = i],$$

which completes the proof of Theorem 3.2 after some routine calculations. □

The discrepancy bound in Theorem 3.2 is not tight. In follow-up work (see Section 5.1), the author proved a substantially stronger bound using matrix-analytic techniques. However, that matrix-analytic approach does not seem to extend to the multiparty model, and as we will see later in Sections 6 and 7, all multiparty papers in this survey use adaptations of the analysis just presented.

## 4 The Generalized Discrepancy Method

As we saw in Section 2.2, the discrepancy method is particularly strong in that it gives communication lower bounds not only for bounded-error protocols but also for protocols with error vanishingly close to $\frac{1}{2}$. Ironically, this strength of the discrepancy method is also its weakness. For example, the DISJOINTNESS function DISJ$(x, y) = \bigvee_{i=1}^{n}(x_i \wedge y_i)$ has a simple low-cost protocol with error $\frac{1}{2} - \Omega\left(\frac{1}{n}\right)$.

As a result, DISJOINTNESS has high discrepancy, and no useful lower bounds can be obtained for it via the discrepancy method. Yet it is well-known that DISJOINTNESS has bounded-error communication complexity $\Omega(n)$ in the randomized model [KS92, Raz92] and $\Omega(\sqrt{n})$ in the quantum model [Raz03].

The remainder of this survey (Sections 5–7) is concerned with bounded-error communication. Crucial to this development is the *generalized* discrepancy method, an ingenious extension of the traditional discrepancy method that avoids the difficulty just cited. To our knowledge, this idea originated in a paper by Klauck [Kla01, Thm. 4] and was reformulated in its current form by Razborov [Raz03]. The development in [Kla01] and [Raz03] takes place in the quantum model of communication. However, the basic mathematical technique is in no way restricted to the quantum model, and we will focus here on a model-independent version of the generalized discrepancy method from [She07b, §2.4].

Specifically, consider an *arbitrary* communication model and let $f : X \times Y \to \{0, 1\}$ be a given function whose communication complexity we wish to estimate. Suppose we can find a function $h : X \times Y \to \{0, 1\}$ and a distribution $\mu$ on $X \times Y$ that satisfy the following two properties.

1. **Correlation of $f$ and $h$.** The functions $f$ and $h$ are well correlated under $\mu$:
$$\mathop{\mathbf{E}}_{(x,y)\sim\mu}\left[(-1)^{f(x,y)+h(x,y)}\right] \geqslant \epsilon, \qquad (4.1)$$

   where $\epsilon > 0$ is typically a constant.

2. **Hardness of $h$.** No low-cost protocol $\Pi$ in the given model of communication can compute $h$ to a substantial advantage under $\mu$. Formally, if $\Pi$ is a protocol in the given model with cost $C$, then
$$\mathop{\mathbf{E}}_{(x,y)\sim\mu}\left[(-1)^{h(x,y)} \mathbf{E}\left[(-1)^{\Pi(x,y)}\right]\right] \leqslant 2^{O(C)}\gamma, \qquad (4.2)$$

   where $\gamma = o(1)$. The inner expectation in (4.2) is over the internal operation of the protocol on the fixed input $(x, y)$.

If the above two conditions hold, we claim that any protocol in the given model that computes $f$ with error at most $\epsilon/3$ on each input must have cost $\Omega\left(\log \frac{\epsilon}{\gamma}\right)$. Indeed, let $\Pi$ be a protocol with $\mathbf{P}[\Pi(x, y) \neq f(x, y)] \leqslant \epsilon/3$ for all $x, y$. Then standard manipulations reveal:

$$\mathop{\mathbf{E}}_{(x,y)\sim\mu}\left[(-1)^{h(x,y)} \mathbf{E}\left[(-1)^{\Pi(x,y)}\right]\right] \geqslant \mathop{\mathbf{E}}_{(x,y)\sim\mu}\left[(-1)^{f(x,y)+h(x,y)}\right] - 2 \cdot \frac{\epsilon}{3} \overset{(4.1)}{\geqslant} \frac{\epsilon}{3}.$$

In view of (4.2), this shows that $\Pi$ must have cost $\Omega\left(\log \frac{\epsilon}{\gamma}\right)$.

The above framework from [She07b] is meant to emphasize the basic mathematical technique in question, which is independent of the communication model. Indeed, the communication model enters the picture only in (4.2). It is here that the analysis must exploit the particularities of the model. To place an upper bound on the advantage under $\mu$ in the quantum model with entanglement, one considers the quantity $\|K\|\sqrt{|X||Y|}$, where $K = [(-1)^{h(x,y)}\mu(x,y)]_{x,y}$. In the randomized model and the quantum model without entanglement, the quantity to estimate happens to be $\mathrm{disc}_\mu(h)$. (In fact, Linial and Shraibman [LS07b] recently showed that $\mathrm{disc}_\mu(h)$ also works in the quantum model with entanglement.)

For future reference, we now record a quantitative version of the generalized discrepancy method for the quantum model.

**Theorem 4.1** ([She07b], implicit in [Raz03, SZ07]). *Let $X, Y$ be finite sets and $f : X \times Y \to \{0, 1\}$ a given function. Let $K = [K_{xy}]_{x \in X, y \in Y}$ be any real matrix with $\|K\|_1 = 1$. Then for each $\epsilon > 0$,*

$$4^{Q_\epsilon(f)} \geqslant 4^{Q^*_\epsilon(f)} \geqslant \frac{\langle F, K \rangle - 2\epsilon}{3\|K\|\sqrt{|X||Y|}},$$

*where $F = \left[(-1)^{f(x,y)}\right]_{x \in X, y \in Y}$.*

Observe that Theorem 4.1 uses slightly more succinct notation (matrix vs. function; weighted sum vs. expectation) but is equivalent to the abstract formulation above.

So far, we have focused on two-party communication. This discussion extends essentially word-for-word to the multiparty model, with discrepancy serving once again as the natural measure of the advantage attainable by low-cost protocols. This extension was formalized by Lee and Shraibman [LS07, Thms. 6, 7] and independently by Chattopadhyay and Ada [CA08, Lem. 3.2], who proved (4.3) and (4.4) below, respectively:

**Theorem 4.2** (cf. [LS07, CA08]). *Fix $F : X_1 \times \cdots \times X_k \to \{-1, +1\}$ and $\epsilon \in [0, 1/2)$. Then*

$$2^{R^k_\epsilon(F)} \geqslant (1 - \epsilon) \max_{H,P} \left\{ \frac{\langle H \circ P, F \rangle - \frac{1}{1-\epsilon}\epsilon}{\mathrm{disc}_P(H)} \right\} \tag{4.3}$$

*and*

$$2^{R^k_\epsilon(F)} \geqslant \max_{H,P} \left\{ \frac{\langle H \circ P, F \rangle - 2\epsilon}{\mathrm{disc}_P(H)} \right\}, \tag{4.4}$$

*where in both cases $H$ ranges over sign tensors and $P$ ranges over tensors with $P \geqslant 0$ and $\|P\|_1 = 1$.*

*Proof.* Fix an optimal $\epsilon$-error protocol $\Pi$ for $F$. Define $\tilde{F}(x_1, \ldots, x_k) = \mathbf{E}[(-1)^{\Pi(x_1,\ldots,x_k)}]$, where the expectation is over any internal randomization in $\Pi$. Let $\delta \in (0, 1]$ be a parameter to be fixed later. Then

$$2^{R^k_\epsilon(F)} \operatorname{disc}_P(H) \geq \langle H \circ P, \tilde{F} \rangle$$

$$= \delta \left\{ \langle H \circ P, F \rangle + \left\langle H \circ P, \frac{1}{\delta}\tilde{F} - F \right\rangle \right\}$$

$$\geq \delta \left\{ \langle H \circ P, F \rangle - \frac{1}{\delta} \max\{|1 - \delta - 2\epsilon|, 1 - \delta\} \right\}.$$

where the first inequality restates the original discrepancy method (Theorem 2.3). Now (4.3) and (4.4) follow by setting $\delta = 1 - \epsilon$ and $\delta = 1$, respectively. □

The proof in [CA08] is similar to the one just given for the special case $\delta = 1$. The proof in [LS07] is rather different and works by defining a suitable norm and passing to its dual. The norm-based approach was employed earlier by Linial and Shraibman [LS07b] and can be thought of as a purely analytic analogue of the generalized discrepancy method.

## 5 Two-Party Bounded-Error Communication

For a function $f : \{0, 1\}^n \to \mathbb{R}$, recall from Section 1 that its $\epsilon$-approximate degree $\deg_\epsilon(f)$ is the least degree of a polynomial $p(x_1, \ldots, x_n)$ with $|f(x) - p(x)| \leq \epsilon$ for all $x \in \{0, 1\}^n$. We move on to discuss two recent papers on bounded-error communication that use the notion of approximate degree to contribute strong lower bounds for rather broad classes of functions, subsuming Razborov's breakthrough work on symmetric predicates [Raz03]. These lower bounds are valid not only in the randomized model, but also in the quantum model (regardless of entanglement).

The setting in which to view these two works is Klauck and Razborov's generalized discrepancy method (see Sections 1 and 4). Let $F$ be a sign matrix whose bounded-error quantum communication complexity is of interest. The quantum version of this method (Theorem 4.1) states that to prove a communication lower bound for $F$, it suffices to exhibit a real matrix $K$ such that $\langle F, K \rangle$ is large but $\|K\|$ is small. The importance of the generalized discrepancy method is that it makes it possible, in theory, to prove lower bounds for functions such as DISJOINTNESS, to which the traditional discrepancy method (Theorem 2.3) does not apply.

The hard part, of course, is finding the matrix $K$. Except in rather restricted cases [Kla01, Thm. 4], it was not known how to do it. As a result, the generalized discrepancy method was of limited practical use. (In particular, Razborov's celebrated work [Raz03] did not use the generalized discrepancy method. Instead, he introduced a novel alternate technique that was restricted to symmetric functions.) This difficulty was overcome independently by Sherstov [She07b] and Shi and Zhu [SZ07], who used the dual characterization of the approximate degree to obtain the matrix $K$ for a broad range of problems. To our knowledge, the work in [She07b] and [SZ07] is the first use of the dual characterization of the approximate degree to prove communication lower bounds.

The specifics of these two works are very different. The construction of $K$ in [She07b], which we called the *pattern matrix method* for lower bounds on bounded-error communication, is built around a new matrix-analytic technique (the *pattern matrix*) inspired by the author's Degree/Discrepancy Theorem. The construction of $K$ in [SZ07], the *block-composition method,* is based on the idea of hardness amplification by composition. What unites them is use of the dual characterization of the approximate degree, given by the following theorem.

**Theorem 5.1** ([She07b, SZ07]). *Fix $\epsilon \geq 0$. Let $f : \{0, 1\}^n \to \mathbb{R}$ be given with $d = \deg_\epsilon(f) \geq 1$. Then there is a function $\psi : \{0, 1\}^n \to \mathbb{R}$ such that:*

$$\hat{\psi}(S) = 0 \qquad \text{for } |S| < d,$$
$$\sum_{z \in \{0,1\}^n} |\psi(z)| = 1,$$
$$\sum_{z \in \{0,1\}^n} \psi(z) f(z) > \epsilon.$$

Theorem 5.1 follows from linear-programming duality. We shall first cover the two papers individually in Sections 5.1 and 5.2 and then compare them in detail in Section 5.3.

## 5.1 The Pattern Matrix Method

The setting for this work resembles that of the Degree/Discrepancy Theorem in [She07a] (see Section 3). Let $N$ and $n$ be positive integers, where $n \leq N/2$. For convenience, we will further assume that $n \mid N$. Fix an arbitrary function $f : \{0, 1\}^n \to \{0, 1\}$. Consider the communication problem of computing

$$f(x|_V),$$

where the bit string $x \in \{0, 1\}^N$ is Alice's input and the set $V \subset \{1, 2, \ldots, N\}$ with $|V| = n$ is Bob's input. As before, $x|_V$ denotes the projection of $x$ onto the indices in $V$, i.e., $x|_V = (x_{i_1}, x_{i_2}, \ldots, x_{i_n}) \in \{0, 1\}^n$ where $i_1 < i_2 < \cdots < i_n$ are the elements of $V$.

The similarities with [She07a], however, do not extend beyond this point. Unlike that earlier work, we will actually study the *easier* communication problem in which Bob's input $V$ is restricted to a rather special form. Namely, we will only allow those sets $V$ that contain precisely one element from each block in the following partition of $\{1, 2, \ldots, N\}$:

$$\left\{1, 2, \ldots, \frac{N}{n}\right\} \cup \left\{\frac{N}{n} + 1, \ldots, \frac{2N}{n}\right\} \cup \cdots \cup \left\{\frac{(n-1)N}{n} + 1, \ldots, N\right\}. \tag{5.1}$$

Even for this easier communication problem, we will prove a much stronger result than what would have been possible in the original setting with the methods of [She07a]. In particular, we will considerably improve the Degree/Discrepancy Theorem from [She07a] along the way. The main results of this work are as follows.

**Theorem 5.2** ([She07b]). *Any classical or quantum protocol, with or without prior entanglement, that computes $f(x|_V)$ with error probability at most $1/5$ on each input has communication cost at least*

$$\frac{1}{4} \deg_{1/3}(f) \cdot \log\left\lfloor \frac{N}{2n} \right\rfloor - 2.$$

In view of the restricted form of Bob's inputs, we can restate Theorem 5.2 in terms of function composition. Setting $N = 4n$ for concreteness, we have:

**Corollary 5.3** ([She07b]). *Let $f : \{0, 1\}^n \to \{0, 1\}$ be given. Define $F : \{0, 1\}^{4n} \times \{0, 1\}^{4n} \to \{0, 1\}$ by*

$$F(x, y) = f\big(\; x_1 y_1 \;\vee\; x_2 y_2 \;\vee\; x_3 y_3 \;\vee\; x_4 y_4 ,$$
$$x_5 y_5 \;\vee\; x_6 y_6 \;\vee\; x_7 y_7 \;\vee\; x_8 y_8 ,$$
$$\vdots$$
$$x_{4n-3} y_{4n-3} \;\vee\; x_{4n-2} y_{4n-2} \;\vee\; x_{4n-1} y_{4n-1} \;\vee\; x_{4n} y_{4n} \;\big),$$

*where $x_i y_i = (x_i \wedge y_i)$. Any classical or quantum protocol, with or without prior entanglement, that computes $F(x, y)$ with error probability at most $1/5$ on each input has cost at least $\frac{1}{4} \deg_{1/3}(f) - 2$.*

We now turn to the proof. Let $\mathcal{V}(N, n)$ denote the set of Bob's inputs, i.e., the family of subsets $V \subseteq [N]$ that have exactly one element in each of the blocks of the partition (5.1). Clearly, $|\mathcal{V}(N, n)| = (N/n)^n$. We will be working with the following family of matrices.

**Definition 5.4** (Pattern matrix [She07b])**.** For $\phi : \{0,1\}^n \to \mathbb{R}$, the $(N, n, \phi)$-*pattern matrix* is the real matrix $A$ given by

$$A = \Big[\phi(x|_V \oplus w)\Big]_{x \in \{0,1\}^N, (V,w) \in \mathcal{V}(N,n) \times \{0,1\}^n}.$$

In words, $A$ is the matrix of size $2^N$ by $2^n(N/n)^n$ whose rows are indexed by strings $x \in \{0,1\}^N$, whose columns are indexed by pairs $(V, w) \in \mathcal{V}(N, n) \times \{0,1\}^n$, and whose entries are given by $A_{x,(V,w)} = \phi(x|_V \oplus w)$. The logic behind the term "pattern matrix" is as follows: a mosaic arises from repetitions of a pattern in the same way that $A$ arises from applications of $\phi$ to various subsets of the variables.

Our intermediate goal will be to determine the spectral norm of any given pattern matrix $A$. Toward that end, we will actually end up determining every singular value of $A$ and its multiplicity. Our approach will be to represent $A$ as the sum of simpler matrices and analyze them instead. For this to work, we need to be able to reconstruct the singular values of $A$ from those of the simpler matrices. Just when this can be done is the subject of the following lemma from [She07b].

**Lemma 5.5** (Singular values of a matrix sum [She07b])**.** *Let $A, B$ be real matrices with $AB^\mathsf{T} = 0$ and $A^\mathsf{T} B = 0$. Then the nonzero singular values of $A + B$, counting multiplicities, are $\sigma_1(A), \ldots, \sigma_{\mathrm{rank}\,A}(A), \sigma_1(B), \ldots, \sigma_{\mathrm{rank}\,B}(B)$.*

We are ready to analyze the singular values of a pattern matrix.

**Theorem 5.6** (Singular values of a pattern matrix [She07b])**.** *Let $\phi : \{0,1\}^n \to \mathbb{R}$ be given. Let $A$ be the $(N, n, \phi)$-pattern matrix. Then the nonzero singular values of $A$, counting multiplicities, are:*

$$\bigcup_{S : \hat{\phi}(S) \neq 0} \left\{ \sqrt{2^{N+n}\left(\frac{N}{n}\right)^n} \cdot |\hat{\phi}(S)| \left(\frac{n}{N}\right)^{|S|/2}, \quad \text{repeated } \left(\frac{N}{n}\right)^{|S|} \text{ times} \right\}.$$

*In particular,*

$$\|A\| = \sqrt{2^{N+n}\left(\frac{N}{n}\right)^n} \max_{S \subseteq [n]} \left\{ |\hat{\phi}(S)| \left(\frac{n}{N}\right)^{|S|/2} \right\}.$$

*Proof* [She07b]. For each $S \subseteq [n]$, let $A_S$ be the $(N, n, \chi_S)$-pattern matrix. Then $A = \sum_{S \subseteq [n]} \hat{\phi}(S) A_S$. For any $S, T \subseteq [n]$ with $S \neq T$, a calculation reveals that $A_S A_T^\mathsf{T} = 0$ and $A_S^\mathsf{T} A_T = 0$. By Lemma 5.5, this means that the nonzero singular values of $A$ are the union of the nonzero singular values of all $\hat{\phi}(S) A_S$, counting multiplicities. Therefore, the proof will be complete once we show that the only nonzero singular value of $A_S^\mathsf{T} A_S$ is $2^{N+n}(N/n)^{n-|S|}$, with multiplicity $(N/n)^{|S|}$.

For this, it is convenient to write $A_S^\mathsf{T} A_S$ as the Kronecker product

$$A_S^\mathsf{T} A_S \;=\; [\chi_S(w)\chi_S(w')]_{w,w'} \;\otimes\; \left[\sum_{x\in\{0,1\}^N} \chi_S(x|_V)\,\chi_S(x|_{V'})\right]_{V,V'}.$$

The first matrix in this factorization has rank 1 and entries $\pm 1$, which means that its only nonzero singular value is $2^n$ with multiplicity 1. The other matrix, call it $M$, is permutation-similar to $2^N \operatorname{diag}(J, J, \ldots, J)$, where $J$ is the all-ones square matrix of order $(N/n)^{n-|S|}$. This means that the only nonzero singular value of $M$ is $2^N(N/n)^{n-|S|}$ with multiplicity $(N/n)^{|S|}$. It follows from elementary properties of the Kronecker product that the spectrum of $A_S^\mathsf{T} A_S$ is as desired. $\square$

We are now prepared to formulate and prove the *pattern matrix method* for lower bounds on bounded-error communication, which gives strong lower bounds for every pattern matrix generated by a Boolean function with high approximate degree. Theorem 5.2 and its corollary will fall out readily as consequences.

**Theorem 5.7** (Pattern matrix method [She07b])**.** *Let $F$ be the $(N, n, f)$-pattern matrix, where $f : \{0,1\}^n \to \{0,1\}$ is given. Put $d = \deg_{1/3}(f)$. Then*

$$Q_{1/5}(F) \;\geqslant\; Q_{1/5}^*(F) \;>\; \frac{1}{4} d \log\left(\frac{N}{n}\right) - 2.$$

*Proof* [She07b]. Define $f^* : \{0,1\}^n \to \{-1, +1\}$ by $f^*(z) = (-1)^{f(z)}$. Then it is easy to verify that $\deg_{2/3}(f^*) = d$. By Theorem 5.1, there is a function $\psi : \{0,1\}^n \to \mathbb{R}$ such that:

$$\hat{\psi}(S) = 0 \qquad\qquad \text{for } |S| < d, \qquad\qquad (5.2)$$

$$\sum_{z\in\{0,1\}^n} |\psi(z)| = 1, \qquad\qquad (5.3)$$

$$\sum_{z\in\{0,1\}^n} \psi(z) f^*(z) > \frac{2}{3}. \qquad\qquad (5.4)$$

Let $M$ be the $(N, n, f^*)$-pattern matrix. Let $K$ be the $(N, n, 2^{-N}(N/n)^{-n}\psi)$-pattern matrix. Immediate consequences of (5.3) and (5.4) are:

$$\|K\|_1 = 1, \qquad \langle K, M \rangle > \frac{2}{3}. \qquad\qquad (5.5)$$

Our last task is to calculate $\|K\|$. By (5.3) and Proposition 2.1,

$$\max_{S \subseteq [n]} |\hat{\psi}(S)| \;\leqslant\; 2^{-n}. \qquad\qquad (5.6)$$

Theorem 5.6 yields, in view of (5.2) and (5.6):

$$\|K\| \leq \left(\frac{n}{N}\right)^{d/2} \left(2^{N+n} \left(\frac{N}{n}\right)^n\right)^{-1/2}. \tag{5.7}$$

The desired lower bounds on quantum communication now follow directly from (5.5) and (5.7) by the generalized discrepancy method (Theorem 4.1). □

*Remark* 5.8. In the proof of Theorem 5.7, we bounded $\|K\|$ using the subtle calculations of the spectrum of a pattern matrix. Another possibility would be to bound $\|K\|$ precisely in the same way that we bounded the discrepancy in the Degree/Discrepancy Theorem (see Section 3). This, however, would result in polynomially weaker lower bounds on communication.

Theorem 5.7 immediately implies Theorem 5.2 above and its corollary:

*Proof of Theorem* 5.2 [She07b]. The $\left(\lfloor \frac{N}{2n} \rfloor n, n, f\right)$-pattern matrix occurs as a submatrix of $[f(x|_V)]_{x \in \{0,1\}^N, V \in \mathcal{V}(N,n)}$. □

**Improved Degree/Discrepancy Theorem.** We will mention a few more applications of this work. The first of these is an improved version of the author's Degree/Discrepancy Theorem (Theorem 3.2).

**Theorem 5.9** ([She07b]). *Let $F$ be the $(N, n, f)$-pattern matrix, where $f : \{0,1\}^n \to \{0,1\}$ has threshold degree $d$. Then $\operatorname{disc}(F) \leq (n/N)^{d/2}$.*

The proof is similar to the proof of the pattern matrix method. Theorem 5.9 improves considerably on the original Degree/Discrepancy Theorem. To illustrate, consider $f(x) = \bigvee_{i=1}^{m} \bigwedge_{j=1}^{m^2} x_{ij}$, a function on $n = m^3$ variables. Applying Theorem 5.9 to $f$ leads to an $\exp(-\Theta(n^{1/3}))$ upper bound on the discrepancy of $\mathsf{AC}^0$, improving on the previous bound of $\exp(-\Theta(n^{1/5}))$ from [She07a]. The $\exp(-\Theta(n^{1/3}))$ bound is also the bound obtained by Buhrman et al. [BVW07] independently of the author [She07a, She07b], using a different function and different techniques.

**Razborov's Lower Bounds for Symmetric Functions.** As another application, we are able to give an alternate proof of Razborov's breakthrough result on the quantum communication complexity of symmetric functions [Raz03]. Consider a communication problem in which Alice has a string $x \in \{0,1\}^n$, Bob has a string $y \in \{0,1\}^n$, and their objective is to compute

$$D(|x \wedge y|)$$

for some predicate $D : \{0, 1, \ldots, n\} \to \{0, 1\}$ fixed in advance. This general setting encompasses several familiar functions, such as DISJOINTNESS (determining if $x$ and $y$ intersect) and INNER PRODUCT MODULO $2$ (determining if $x$ and $y$ intersect in an odd number of positions).

As it turns out, the hardness of this general communication problem depends on whether $D$ changes value close to the middle of the range $\{0, 1, \ldots, n\}$. Specifically, define $\ell_0(D) \in \{0, 1, \ldots, \lfloor n/2 \rfloor\}$ and $\ell_1(D) \in \{0, 1, \ldots, \lceil n/2 \rceil\}$ to be the smallest integers such that $D$ is constant in the range $[\ell_0(D), n - \ell_1(D)]$. Razborov established optimal lower bounds on the quantum communication complexity of every function of the form $D(|x \wedge y|)$:

**Theorem 5.10** (Razborov [Raz03])**.** *Let* $D : \{0, 1, \ldots, n\} \to \{0, 1\}$ *be an arbitrary predicate. Put* $f(x, y) = D(|x \wedge y|)$. *Then*

$$Q_{1/3}(f) \geq Q^*_{1/3}(f) \geq \Omega\left(\sqrt{n\ell_0(D)} + \ell_1(D)\right).$$

In particular, DISJOINTNESS has quantum communication complexity $\Omega(\sqrt{n})$, regardless of entanglement. Prior to Razborov's result, the best lower bound [BW01, ASTS$^+$03] for DISJOINTNESS was only $\Omega(\log n)$.

In [She07b], we give a new proof of Razborov's Theorem 5.10 using a straightforward application of the pattern matrix method.

## 5.2 The Block Composition Method

Given functions $f : \{0, 1\}^n \to \{0, 1\}$ and $g : \{0, 1\}^k \times \{0, 1\}^k \to \{0, 1\}$, let $f \circ g^n$ denote the composition of $f$ with $n$ independent copies of $g$. More formally, the function $f \circ g^n : \{0, 1\}^{nk} \times \{0, 1\}^{nk} \to \{0, 1\}$ is given by

$$(f \circ g^n)(x, y) = f(\ldots, g(x^{(i)}, y^{(i)}), \ldots),$$

where $x = (\ldots, x^{(i)}, \ldots) \in \{0, 1\}^{nk}$ and $y = (\ldots, y^{(i)}, \ldots) \in \{0, 1\}^{nk}$.

This section presents Shi and Zhu's *block composition method* [SZ07], which gives a lower bound on the communication complexity of $f \circ g^n$ in terms of certain properties of $f$ and $g$. The relevant property of $f$ is simply its approximate degree. The relevant property of $g$ is its *spectral discrepancy,* formalized next.

**Definition 5.11** (Spectral discrepancy [SZ07])**.** *Given* $g : \{0, 1\}^k \times \{0, 1\}^k \to \{0, 1\}$, *its spectral discrepancy* $\rho(g)$ *is the least* $\rho \geq 0$ *for which there exist sets* $A, B \subseteq$

$\{0,1\}^k$ and a distribution $\mu$ on $A \times B$ such that

$$\left\| \left[ \mu(x,y)(-1)^{g(x,y)} \right]_{x \in A, y \in B} \right\| \leq \frac{\rho}{\sqrt{|A|\,|B|}}, \tag{5.8}$$

$$\left\| \left[ \mu(x,y) \right]_{x \in A, y \in B} \right\| \leq \frac{1+\rho}{\sqrt{|A|\,|B|}}, \tag{5.9}$$

and

$$\sum_{(x,y) \in A \times B} \mu(x,y)(-1)^{g(x,y)} = 0. \tag{5.10}$$

In view of (5.8) alone, the spectral discrepancy $\rho(g)$ is an upper bound on the discrepancy $\mathrm{disc}(g)$. The key additional requirement (5.9) is satisfied, for example, by doubly stochastic matrices [HJ86, §8.7]: if $A = B$ and all row and column sums in $[\mu(x,y)]_{x \in A, y \in A}$ are $1/|A|$, then $\|[\mu(x,y)]_{x \in A, y \in A}\| = 1/|A|$.

As an illustration, consider the familiar function INNER PRODUCT MODULO 2, given by $\mathrm{IP}_k(x,y) = \bigoplus_{i=1}^{k}(x_i \wedge y_i)$.

**Proposition 5.12** ([SZ07]). *The function* $\mathrm{IP}_k$ *has* $\rho(\mathrm{IP}_k) \leq 1/\sqrt{2^k - 1}$.

*Proof* [SZ07]. Take $\mu$ to be the uniform distribution over $A \times B$, where $A = \{0,1\}^k \setminus \{0^k\}$ and $B = \{0,1\}^k$. □

We are prepared to state the general method.

**Theorem 5.13** (Block composition method [SZ07]). *Fix* $f : \{0,1\}^n \to \{0,1\}$ *and* $g : \{0,1\}^k \times \{0,1\}^k \to \{0,1\}$. *Put* $d = \deg_{1/3}(f)$ *and* $\rho = \rho(g)$. *If* $\rho \leq d/(2en)$, *then*

$$Q(f \circ g^n) \geq Q^*(f \circ g^n) = \Omega(d).$$

*Proof* (adapted from [SZ07]). Fix sets $A, B \subseteq \{0,1\}^k$ and a distribution $\mu$ on $A \times B$ with respect to which $\rho = \rho(g)$ is achieved. Define $f^* : \{0,1\}^n \to \{-1,+1\}$ by $f^*(z) = (-1)^{f(z)}$. Then one readily verifies that $\deg_{2/3}(f^*) = d$. By Theorem 5.1, there exists $\psi : \{0,1\}^n \to \mathbb{R}$ such that

$$\hat{\psi}(S) = 0 \qquad \text{for } |S| < d, \tag{5.11}$$

$$\sum_{z \in \{0,1\}^n} |\psi(z)| = 1, \tag{5.12}$$

$$\sum_{z \in \{0,1\}^n} \psi(z) f^*(z) > \frac{2}{3}. \tag{5.13}$$

Define matrices

$$F = \left[f^*(\ldots, g(x^{(i)}, y^{(i)}), \ldots)\right]_{x,y},$$

$$K = \left[2^n \psi(\ldots, g(x^{(i)}, y^{(i)}), \ldots) \prod_{i=1}^{n} \mu(x^{(i)}, y^{(i)})\right]_{x,y},$$

where in both cases the row index $x = (\ldots, x^{(i)}, \ldots)$ ranges over $A^n$ and the column index $y = (\ldots, y^{(i)}, \ldots)$ ranges over $B^n$. In view of (5.10) and (5.13),

$$\langle F, K \rangle > \frac{2}{3}. \tag{5.14}$$

We proceed to bound $\|K\|$. Put

$$M_S = \left[\prod_{i \in S}(-1)^{g(x^{(i)}, y^{(i)})} \cdot \prod_{i=1}^{n} \mu(x^{(i)}, y^{(i)})\right]_{x,y}, \qquad S \subseteq [n].$$

Then (5.8) and (5.9) imply, in view of the tensor structure of $M_S$, that

$$\|M_S\| \leqslant |A|^{-n/2} |B|^{-n/2} \rho^{|S|}(1 + \rho)^{n-|S|}. \tag{5.15}$$

On the other hand,

$$\begin{aligned}
\|K\| &\leqslant \sum_{S \subseteq [n]} 2^n |\hat{\psi}(S)| \|M_S\| \\
&= \sum_{|S| \geqslant d} 2^n |\hat{\psi}(S)| \|M_S\| && \text{by (5.11)} \\
&\leqslant \sum_{|S| \geqslant d} \|M_S\| && \text{by (5.12) and Proposition 2.1} \\
&\leqslant |A|^{-n/2} |B|^{-n/2} \sum_{i=d}^{n} \binom{n}{i} \rho^i (1 + \rho)^{n-i} && \text{by (5.15).}
\end{aligned}$$

Since $\rho \leqslant d/(2en)$, we further have

$$\|K\| \leqslant |A|^{-n/2} |B|^{-n/2} 2^{-\Theta(d)}. \tag{5.16}$$

In view of (5.14) and (5.16), the desired lower bound on $Q^*(F)$ now follows by the generalized discrepancy method (Theorem 4.1). □

Proposition 5.12 and Theorem 5.13 have the following consequence:

**Theorem 5.14** ([SZ07]). *Fix a function* $f : \{0, 1\}^n \to \{0, 1\}$ *and an integer* $k \geq 2\log_2 n + 5$. *Then* $Q(f \circ \text{IP}_k^n) \geq Q^*(f \circ \text{IP}_k^n) \geq \Omega(\deg_{1/3}(f))$.

For the DISJOINTNESS function $\text{DISJ}_k(x, y) = \bigvee_{i=1}^{k}(x_i \wedge y_i)$, Shi and Zhu prove that $\rho(\text{DISJ}_k) = O(1/k)$. Unlike Proposition 5.12, this fact requires a nontrivial proof using Knuth's calculation of the eigenvalues of certain combinatorial matrices. In conjunction with Theorem 5.13, this upper bound on $\rho(\text{DISJ}_k)$ leads with some work to the following implication:

**Theorem 5.15** ([SZ07]). *Define* $f : \{0, 1\}^n \times \{0, 1\}^n \to \{0, 1\}$ *by* $f(x, y) = D(|x \wedge y|)$, *where* $D : \{0, 1, \ldots, n\} \to \{0, 1\}$ *is given. Then*

$$Q(f) \geq Q^*(f) \geq \Omega\left(n^{1/3} \ell_0(D)^{2/3} + \ell_1(D)\right).$$

The symbols $\ell_0(D)$ and $\ell_1(D)$ have their meaning from Section 5.1. Theorem 5.15 is of course a weaker version of Razborov's celebrated lower bounds for symmetric functions (Theorem 5.10), obtained with a different proof.

### 5.3 Pattern Matrix Method vs. Block Composition Method

To restate the block composition method,

$$Q^*(f \circ g^n) \geq \Omega(\deg_{1/3}(f)) \qquad \text{provided that} \qquad \rho(g) \leq \frac{\deg_{1/3}(f)}{2en}.$$

The key player in this method is the quantity $\rho(g)$, which needs to be small. This poses two complications. First, the function $g$ will generally need to depend on many variables, from $k = \Theta(\log n)$ to $k = n^{\Theta(1)}$, which weakens the final lower bounds on communication (recall that $\rho(g) \geq 2^{-k}$ always). For example, the lower bounds obtained in [SZ07] for symmetric functions are polynomially weaker than Razborov's optimal lower bounds (see Theorems 5.15 and 5.10, respectively).

A second complication, as Shi and Zhu note, is that "estimating the quantity $\rho(g)$ is unfortunately difficult in general" [SZ07, §4.1]. For example, re-proving Razborov's lower bounds reduces to estimating $\rho(g)$ with $g$ being the DISJOINTNESS function. Shi and Zhu accomplish this using Hahn matrices, an advanced tool that is also the centerpiece of Razborov's own proof (Razborov's use of Hahn matrices is somewhat more demanding).

These complications do not arise in the pattern matrix method. For example, it implies (by setting $N = 2n$ in Theorem 5.7) that

$$Q^*(f \circ g^n) \geq \Omega(\deg_{1/3}(f))$$

for any function $g : \{0, 1\}^k \times \{0, 1\}^k \to \{0, 1\}$ such that the matrix $[g(x, y)]_{x,y}$ contains the following submatrix, up to permutations of rows and columns:

$$\begin{bmatrix} 1 & 0 & 1 & 0 \\ 1 & 0 & 0 & 1 \\ 0 & 1 & 1 & 0 \\ 0 & 1 & 0 & 1 \end{bmatrix}. \tag{5.17}$$

To illustrate, one can take $g$ to be

$$g(x, y) = x_1 y_1 \vee x_2 y_2 \vee x_3 y_3 \vee x_4 y_4,$$

or

$$g(x, y) = x_1 y_1 y_2 \vee \overline{x_1}\, y_1 \overline{y_2} \vee x_2 \overline{y_1}\, y_2 \vee \overline{x_2}\, \overline{y_1}\, \overline{y_2}.$$

(In particular, the pattern matrix method subsumes Theorem 5.14.) To summarize, there is a simple function $g$ on only $k = 2$ variables that works universally for all $f$. This means no technical conditions to check, such as $\rho(g)$, and no blow-up in the number of variables. As a result, in [She07b] we are able to re-prove Razborov's optimal lower bounds exactly. Moreover, the technical machinery involved is self-contained and disjoint from Razborov's proof.

We have just seen that the pattern matrix method gives strong lower bounds for many functions to which the block composition method does not apply. However, this does not settle the exact relationship between the scopes of applicability of the two methods. Several natural questions arise. If a function $g : \{0, 1\}^k \times \{0, 1\}^k \to \{0, 1\}$ has spectral discrepancy $\rho(g) \leqslant \frac{1}{2e}$, does the matrix $[g(x, y)]_{x,y}$ contain (5.17) as a submatrix, up to permutations of rows and columns? An affirmative answer would mean that the pattern matrix method has a strictly greater scope of applicability; a negative answer would mean that the block composition method works in some situations where the pattern matrix method does not apply. If the answer is negative, what can be said for $\rho(g) = o(1)$ or $\rho(g) = n^{-\Theta(1)}$?

Another intriguing issue concerns multiparty communication. As we will see in Section 6, the pattern matrix method extends readily to the multiparty model. This extension makes heavy use of the fact that the rows of a pattern matrix are applications of the same function to different subsets of the variables. In the general context of block composition (Section 5.2), it is unclear how to carry out this extension. It is inviting to explore a synthesis of the two methods in the multiparty model or another suitable context.

## 6 Extensions to the Multiparty Model

In this section, we present extensions of the Degree/Discrepancy Theorem and of the pattern matrix method to the multiparty model. We start with some notation.

Fix a function $\phi : \{0,1\}^n \to \mathbb{R}$ and an integer $N$ with $n \mid N$. Define the $(k, N, n, \phi)$-*pattern tensor* as the $k$-argument function $A : \{0,1\}^{n(N/n)^{k-1}} \times [N/n]^n \times \cdots \times [N/n]^n \to \mathbb{R}$ given by $A(x, V_1, \ldots, V_{k-1}) = \phi(x|_{V_1,\ldots,V_{k-1}})$, where

$$x|_{V_1,\ldots,V_{k-1}} \stackrel{\text{def}}{=} (x_{1,V_1[1],\ldots,V_{k-1}[1]}, \ldots, x_{n,V_1[n],\ldots,V_{k-1}[n]}) \in \{0,1\}^n$$

and $V_j[i]$ denotes the $i$th element of the $n$-dimensional vector $V_j$. (Note that we index the string $x$ by viewing it as a $k$-dimensional array of $n \times (N/n) \times \cdots \times (N/n) = n(N/n)^{k-1}$ bits.) This definition generalizes the author's pattern matrices if one ignores the $\oplus$ operator (Section 5.1).

We are ready for the first result of this section, namely, an extension of the Degree/Discrepancy Theorem (Theorem 3.2) to the multiparty model. This extension was originally obtained by Chattopadhyay [Cha07, Lem. 2] for slightly different tensors and has since been revisited in one form or another: [LS07, Thm. 19], [CA08, Lem. 4.2]. The proofs of these several versions are quite similar and are in close correspondence with the original two-party case.

**Theorem 6.1** ([Cha07, LS07, CA08]). *Let $f : \{0,1\}^n \to \{0,1\}$ be given with threshold degree $d \geq 1$. Let $N$ be a given integer, $n \mid N$. Let $F$ be the $(k, N, n, f)$-pattern tensor. If $N \geq 4en^2(k-1)2^{2^{k-1}}/d$, then $\mathrm{disc}(F) \leq 2^{-d/2^{k-1}}$.*

*Proof* (adapted from [Cha07, LS07, CA08]). As in the proof of the Degree/Discrepancy Theorem, let $\mu$ be a probability distribution over $\{0,1\}^n$ with respect to which $\mathbf{E}_{z \sim \mu}[(-1)^{f(z)}p(z)] = 0$ for every real-valued function $p$ of $d-1$ or fewer of the variables $z_1, \ldots, z_n$. The existence of $\mu$ is assured by Theorem 3.3. We will analyze the discrepancy of $F$ with respect to the distribution

$$\lambda(x, V_1, \ldots, V_{k-1}) = 2^{-n(N/n)^{k-1}+n} \left(\frac{N}{n}\right)^{-n(k-1)} \mu(x|_{V_1,\ldots,V_{k-1}}).$$

Define $\psi : \{0,1\}^n \to \mathbb{R}$ by $\psi(z) = (-1)^{f(z)}\mu(z)$. By Theorem 2.2,

$$\mathrm{disc}_\lambda(F)^{2^{k-1}} \leq 2^{n2^{k-1}} \mathbf{E}_{\mathbf{V}} |\Gamma(\mathbf{V})|, \tag{6.1}$$

where we put $\mathbf{V} = (V_1^0, V_1^1, \ldots, V_{k-1}^0, V_{k-1}^1)$ and

$$\Gamma(\mathbf{V}) = \mathbf{E}_x \left[ \underbrace{\psi\left(x|_{V_1^0, V_2^0, \ldots, V_{k-1}^0}\right)}_{(\dagger)} \underbrace{\prod_{z \in \{0,1\}^{k-1} \setminus \{0^{k-1}\}} \psi\left(x|_{V_1^{z_1}, V_2^{z_2}, \ldots, V_{k-1}^{z_{k-1}}}\right)}_{(\ddagger)} \right].$$

For a fixed choice of **V**, define sets

$$A = \{(i, V_1^0[i], \ldots, V_{k-1}^0[i]) : i = 1, 2, \ldots, n\},$$
$$B = \{(i, V_1^{z_1}[i], \ldots, V_{k-1}^{z_{k-1}}[i]) : i = 1, 2, \ldots, n; \ z \in \{0, 1\}^{k-1} \setminus \{0^{k-1}\}\}.$$

Clearly, $A$ and $B$ are the sets of variables featured in the expressions (†) and (‡) above, respectively. To analyze $\Gamma(\mathbf{V})$, we prove two key claims analogous to those in the Degree/Discrepancy Theorem.

**Claim 6.2.** *Assume that $|A \cap B| \leq d - 1$. Then $\Gamma(\mathbf{V}) = 0$.*

*Proof.* Immediate from the fact that the Fourier transform of $\psi$ is supported on characters of order $d$ and higher. □

**Claim 6.3.** *Assume that $|A \cap B| = i$. Then $|\Gamma(\mathbf{V})| \leq 2^{i2^{k-1} - n2^{k-1}}$.*

*Proof.* Observation 3.4 shows that $|\Gamma(\mathbf{V})| \leq 2^{-n2^{k-1}} 2^{n2^{k-1} - |A \cup B|}$. Furthermore, it is straightforward to verify that $|A \cup B| \geq n2^{k-1} - |A \cap B| 2^{k-1}$. □

In view of Claims 6.2 and 6.3, inequality (6.1) simplifies to

$$\mathrm{disc}_\lambda(F)^{2^{k-1}} \leq \sum_{i=d}^{n} 2^{i2^{k-1}} \mathbf{P}[|A \cap B| = i].$$

It remains to bound $\mathbf{P}[|A \cap B| = i]$. For a fixed element $a$, we have $\mathbf{P}[a \in B \mid a \in A] \leq (k-1)n/N$ by the union bound. Moreover, given two distinct elements $a, a' \in A$, the corresponding events $a \in B$ and $a' \in B$ are independent. Therefore, $\mathbf{P}[|A \cap B| = i] \leq \binom{n}{i} \left(\frac{(k-1)n}{N}\right)^i$, which yields the desired bound on $\mathrm{disc}_\lambda(F)$. □

*Remark* 6.4. Recall from Section 5.1 that the two-party Degree/Discrepancy Theorem was considerably improved in [She07b] using matrix-analytic techniques. Those techniques, however, do not extend to the multiparty model. As a result, Theorem 6.1 that we have just presented does not subsume the improved Degree/Discrepancy Theorem (Theorem 5.9).

We now present an adaptation of the pattern matrix method (Theorem 5.7) to the multiparty model, obtained by Lee and Shraibman [LS07] and independently by Chattopadhyay and Ada [CA08]. The proof is closely analogous to the two-party case. However, the spectral calculations for pattern matrices do not extend to the multiparty model, and one is forced to fall back on the less precise calculations introduced in the Degree/Discrepancy Theorem (Theorem 3.2). In particular, the result we are about to present does not subsume the two-party pattern matrix method.

**Theorem 6.5** ([LS07, CA08]). *Let $f : \{0, 1\}^n \to \{0, 1\}$ be given with $\deg_{1/3}(f) = d \geq 1$. Let $N$ be a given integer, $n \mid N$. Let $F$ be the $(k, N, n, f)$-pattern tensor. If $N \geq 4en^2(k-1)2^{2^{k-1}}/d$, then $R^k(F) \geq \Omega(d/2^k)$.*

*Proof* (adapted from [LS07, CA08]). Define $f^* : \{0, 1\}^n \to \{-1, +1\}$ by $f^*(z) = (-1)^{f(z)}$. Then it is easy to verify that $\deg_{2/3}(f^*) = d$. By Theorem 5.1, there is a function $\psi : \{0, 1\}^n \to \mathbb{R}$ such that:

$$\hat{\psi}(S) = 0 \qquad \text{for } |S| < d,$$

$$\sum_{z \in \{0,1\}^n} |\psi(z)| = 1,$$

$$\sum_{z \in \{0,1\}^n} \psi(z) f^*(z) > \frac{2}{3}. \tag{6.2}$$

Fix a function $h : \{0, 1\}^n \to \{-1, +1\}$ and a distribution $\mu$ on $\{0, 1\}^n$ such that $\psi(z) \equiv h(x)\mu(x)$. Let $H$ be the $(k, N, n, h)$-pattern tensor. Let $P$ be the $(k, N, n, 2^{-n(N/n)^{k-1}+n}(N/n)^{-n(k-1)}\mu)$-pattern tensor. Then $P$ is a probability distribution. By (6.2),

$$\langle H \circ P, F^* \rangle > \frac{2}{3}, \tag{6.3}$$

where $F^*$ is the $(k, N, n, f^*)$-pattern tensor. As we saw in the proof of Theorem 6.1,

$$\mathrm{disc}_P(H) \leq 2^{-d/2^{k-1}}. \tag{6.4}$$

The theorem now follows by the generalized discrepancy method (Theorem 4.2) in view of (6.3) and (6.4). □

The authors of [LS07] and [CA08] gave important applications of their work to the $k$-party randomized communication complexity of DISJOINTNESS, improving it from $\Omega(\frac{1}{k} \log n)$ to $n^{\Omega(1/k)} 2^{-O(2^k)}$. As a corollary, they separated the multiparty communication classes $\mathsf{NP}^{cc}_k$ and $\mathsf{BPP}^{cc}_k$ for $k = (1 - o(1)) \log_2 \log_2 n$ parties. They also obtained new results for Lovász-Schrijver proof systems, in light of the work due to Beame, Pitassi, and Segerlind [BPS07].

# 7 Separation of $\mathsf{NP}^{cc}_k$ and $\mathsf{BPP}^{cc}_k$

We conclude this survey with a separation of $\mathsf{NP}^{cc}_k$ and $\mathsf{BPP}^{cc}_k$ for $k = (1-\epsilon)\log_2 n$ parties, due to David and Pitassi [DP08]. This is an exponential improvement over the previous separation in [LS07, CA08]. The crucial insight in this new work is to redefine the projection operator $x|_{V_1,...,V_{k-1}}$ from Section 6 using the probabilistic method. This removes the key bottleneck in the previous analyses [LS07, CA08]. Unlike the previous work, however, this new approach no longer applies to DISJOINTNESS.

We start with some notation. Fix integers $n, m$ with $n > m$. Let $\psi : \{0, 1\}^m \to \mathbb{R}$ be a given function with $\sum_{z \in \{0,1\}^m} |\psi(z)| = 1$. Let $d$ denote the least order of a nonzero Fourier coefficient of $\psi$. Fix a Boolean function $h : \{0, 1\}^m \to \{-1, +1\}$ and a distribution $\mu$ on $\{0, 1\}^m$ such that $\psi(z) \equiv h(z)\mu(z)$. For a mapping $\alpha : (\{0, 1\}^n)^k \to \binom{[n]}{m}$, define a $(k + 1)$-party communication problem $H_\alpha : (\{0, 1\}^n)^{k+1} \to \{-1, +1\}$ by $H(x, y_1, \ldots, y_k) = h(x|_{\alpha(y_1,\ldots,y_k)})$. Analogously, define a distribution $\lambda_\alpha$ on $(\{0, 1\}^n)^{k+1}$ by $\lambda(x, y_1, \ldots, y_k) = 2^{-(k+1)n+m} \mu(x|_{\alpha(y_1,\ldots,y_k)})$.

**Theorem 7.1** ([DP08]). *Assume that $n \geq 16em^2 2^k$. Then for a uniformly random choice of $\alpha : (\{0, 1\}^n)^k \to \binom{[n]}{m}$,*

$$\mathop{\mathbf{E}}_\alpha \left[ \mathrm{disc}_{\lambda_\alpha}(H_\alpha)^{2^k} \right] \leq 2^{-n/2} + 2^{-d2^k+1}.$$

*Proof* (adapted from [DP08]). By Theorem 2.2,

$$\mathrm{disc}_{\lambda_\alpha}(H_\alpha)^{2^k} \leq 2^{m2^k} \mathop{\mathbf{E}}_Y |\Gamma(Y)|, \tag{7.1}$$

where we put $Y = (y_1^0, y_1^1, \ldots, y_k^0, y_k^1)$ and

$$\Gamma(Y) = \mathop{\mathbf{E}}_x \left[ \prod_{z \in \{0,1\}^k} \psi\left(x|_{\alpha(y_1^{z_1}, y_2^{z_2}, \ldots, y_k^{z_k})}\right) \right].$$

For a fixed choice of $Y$, we will use the shorthand $S_z = \alpha(y_1^{z_1}, \ldots, y_k^{z_k})$. To analyze $\Gamma(Y)$, we prove two key claims analogous to those in the Degree/Discrepancy Theorem and in Theorem 6.1.

**Claim 7.2.** *Assume that $|\bigcup S_z| > m2^k - d2^{k-1}$. Then $\Gamma(Y) = 0$.*

*Proof.* If $|\bigcup S_z| > m2^k - d2^{k-1}$, then some $S_z$ must feature more than $m - d$ elements that do not occur in $\bigcup_{u \neq z} S_u$. But this forces $\Gamma(Y) = 0$ since the Fourier transform of $\psi$ is supported on characters of order $d$ and higher. □

**Claim 7.3.** *For every $Y$, $|\Gamma(Y)| \leq 2^{-|\bigcup S_z|}$.*

*Proof.* Immediate from Observation 3.4. □

In view of (7.1) and Claims 7.2 and 7.3, we have

$$\mathop{\mathbf{E}}_{\alpha}\left[\mathrm{disc}_{\lambda_\alpha}(H_\alpha)^{2^k}\right] \leq \sum_{i=d2^{k-1}}^{m2^k-m} 2^i \mathop{\mathbf{P}}_{Y,\alpha}\left[\left|\bigcup S_z\right| = m2^k - i\right].$$

It remains to bound the probabilities in the last expression. With probability at least $1 - k2^{-n}$ over the choice of $Y$, the strings $y_1^0, y_1^0 \ldots, y_k^0, y_k^1$ will all be distinct. Conditioning on this event, the fact that $\alpha$ is chosen uniformly at random means that the $2^k$ sets $S_z$ are distributed independently and uniformly over $\binom{[n]}{m}$. A calculation now reveals that

$$\mathop{\mathbf{P}}_{Y,\alpha}\left[\left|\bigcup S_z\right| = m2^k - i\right] \leq k2^{-n} + \binom{m2^k}{i}\left(\frac{m2^k}{n}\right)^i \leq k2^{-n} + 8^{-i}. \qquad \square$$

We are ready to present the separation of $\mathsf{NP}_k^{cc}$ and $\mathsf{BPP}_k^{cc}$.

**Theorem 7.4** (Separation of $\mathsf{NP}_k^{cc}$ and $\mathsf{BPP}_k^{cc}$ [DP08]). *Let $k \leq (1-\epsilon)\log_2 n$, where $\epsilon > 0$ is a given constant. Then there exists a function $F_\alpha : (\{0,1\}^n)^{k+1} \to \{-1,+1\}$ with $N^{k+1}(F_\alpha) = O(\log n)$ but $R^{k+1}(F_\alpha) = n^{\Omega(1)}$.*

*Proof* (adapted from [DP08]). Let $m = \lfloor n^\zeta \rfloor$ for a sufficiently small constant $\zeta = \zeta(\epsilon) > 0$. As usual, define $\mathrm{OR}_m : \{0,1\}^m \to \{-1,+1\}$ by $\mathrm{OR}_m(z) = 1 \Leftrightarrow z = 0^m$. It is known [NS92, Pat92] that $\deg_{1/3}(\mathrm{OR}_m) = \Theta(\sqrt{m})$. As a result, Theorem 5.1 guarantees the existence of a function $\psi : \{0,1\}^m \to \mathbb{R}$ such that:

$$\hat\psi(S) = 0 \qquad\qquad \text{for } |S| < \Theta(\sqrt{m}),$$

$$\sum_{z \in \{0,1\}^m} |\psi(z)| = 1,$$

$$\sum_{z \in \{0,1\}^m} \psi(z) \mathrm{OR}_m(z) > \frac{1}{3}.$$

Fix a function $h : \{0,1\}^m \to \{-1,+1\}$ and a distribution $\mu$ on $\{0,1\}^m$ such that $\psi(z) \equiv h(z)\mu(z)$. For a mapping $\alpha : (\{0,1\}^n)^k \to \binom{[n]}{m}$, let $H_\alpha$ and $\lambda_\alpha$ be as defined at the beginning of this section. Then Theorem 7.1 shows the existence of $\alpha$ such that

$$\mathrm{disc}_{\lambda_\alpha}(H_\alpha) \leq 2^{-\Omega(\sqrt{m})}.$$

Using the properties of $\psi$, one readily verifies that $\langle H \circ \lambda_\alpha, F_\alpha \rangle \geq 1/3$, where $F_\alpha : (\{0,1\}^n)^{k+1} \to \{-1,+1\}$ is given by $F_\alpha(x, y_1, \ldots, y_k) = \mathrm{OR}_m(x|_{\alpha(y_1,\ldots,y_k)})$. By the generalized discrepancy method (Theorem 4.2),

$$R^{k+1}(F_\alpha) \geq \Omega(\sqrt{m}) = n^{\Omega(1)}.$$

On the other hand, $F_\alpha$ has nondeterministic complexity $O(\log n)$. Namely, Player 1 (who knows $y_1, \ldots, y_k$) nondeterministically selects an element $i \in \alpha(y_1, \ldots, y_k)$ and announces $i$. Player 2 (who knows $x$) then announces $x_i$ as the output of the protocol. □

A recent follow-up result due to David, Pitassi, and Viola [DPV08] derandomizes the choice of $\alpha$ in Theorem 7.4, yielding an *explicit* separation of $\mathsf{NP}^{cc}_k$ and $\mathsf{BPP}^{cc}_k$ for $k \leq (1 - \epsilon) \log_2 n$.

# Acknowledgments

I would like to thank Anil Ada, Boaz Barak, Arkadev Chattopadhyay, Adam Klivans, Troy Lee, Yaoyun Shi, and Ronald de Wolf for their helpful feedback on a preliminary version of this survey.